\newcommand{\CLASSINPUTtoptextmargin}{1.905cm}
\newcommand{\CLASSINPUTbottomtextmargin}{4.05cm}

\documentclass[conference, a4paper, 10.9pt]{IEEEtran}
\IEEEoverridecommandlockouts
\usepackage{amsmath,amssymb,amsfonts}
\usepackage{algorithmic}
\usepackage{graphicx}
\usepackage{textcomp}
\usepackage{xcolor}
\def\BibTeX{{\rm B\kern-.05em{\sc i\kern-.025em b}\kern-.08em
    T\kern-.1667em\lower.7ex\hbox{E}\kern-.125emX}}
    
\usepackage[sorting=none]{biblatex}

\usepackage{balance}
\usepackage{csquotes}
\usepackage{comment}
\usepackage{amsmath}    
\usepackage{mathptmx}
\renewcommand{\figurename}{Fig.}
\usepackage[singlespacing]{setspace} 
\begin{document}

\title{Advancing Ischemic Stroke Diagnosis: A Novel Two-Stage Approach for Blood Clot Origin Identification}

\author{\IEEEauthorblockN{Koushik Sivarama Krishnan\IEEEauthorrefmark{1},
Joe Nikesh Puthota John\IEEEauthorrefmark{2},
Swathi Gnanasekar\IEEEauthorrefmark{3},
Karthik Sivarama Krishnan\IEEEauthorrefmark{4},
}
\IEEEauthorblockA{\\
Email: koushik.nov01@gmail.com\IEEEauthorrefmark{1}, joenikesh3@gmail.com\IEEEauthorrefmark{2},
swathisekar727@gmail.com\IEEEauthorrefmark{3},
ks7585@g.rit.edu\IEEEauthorrefmark{4}}}

\maketitle

\begin{abstract}
An innovative two-stage methodology for categorizing blood clot origins is presented in this paper, which is important for the diagnosis and treatment of ischemic stroke. First, a background classifier based on MobileNetV3 segments big whole-slide digital pathology images into numerous tiles to detect the presence of cellular material. After that, different pre-trained image classification algorithms are fine-tuned to determine the origin of blood clots. Due to complex blood flow dynamics and limitations in conventional imaging methods such as computed tomography (CT), magnetic resonance imaging (MRI), and ultrasound, identifying the sources of blood clots is a challenging task. Although these techniques are useful for identifying blood clots, they are not very good at determining how they originated. To address these challenges, our method makes use of robust computer vision models that have been refined using information from whole-slide digital pathology images. Out of all the models tested, the PoolFormer \cite{yu2022metaformer} performs better than the others, with 93.4\% accuracy, 93.4\% precision, 93.4\% recall, and 93.4\% F1-score. Moreover, it achieves the good weighted multi-class logarithmic loss (WMCLL) of 0.4361, which emphasizes how effective it is in this particular application. These encouraging findings suggest that our approach can successfully identify the origin of blood clots in a variety of vascular locations, potentially advancing ischemic stroke diagnosis and treatment approaches.\\
\end{abstract}

\begin{IEEEkeywords}
ischemic stroke; PoolFormer; computer vision; blood clot origin;
\end{IEEEkeywords}

\section{Introduction}
Stroke is considered the leading cause of death and disability globally, with approximately 12.2 million new cases reported annually \cite{worldstrokeorg}. It occurs when the brain's blood supply is interrupted, leading to brain damage and potentially irreversible neurological impairments. Ischemic stroke is the most common type, resulting from a blockage in the blood vessel that supplies blood to the brain. The source of this blockage can be a blood clot originating from various parts of the body such as the heart, carotid artery, or other arteries. Identifying the source of the blood clot is essential for the effective prevention and treatment of stroke. There has been an increasing interest in developing methods to classify the origin of stroke blood clots in recent times. This paper aims to create an intuitive and easy-to-use system that performs well in classifying the origin of blood clots in ischemic stroke patients.\\

The article will commence by highlighting the significance of detecting the source of blood clots that cause strokes. Subsequently, it will analyze the different types of blood clots, including their incidence and features. The article will then delve into the proposed method that can  classify the blood clots of ischemic stroke patients effectively.

\begin{figure}
    \includegraphics[width=0.50\textwidth, height=80mm]{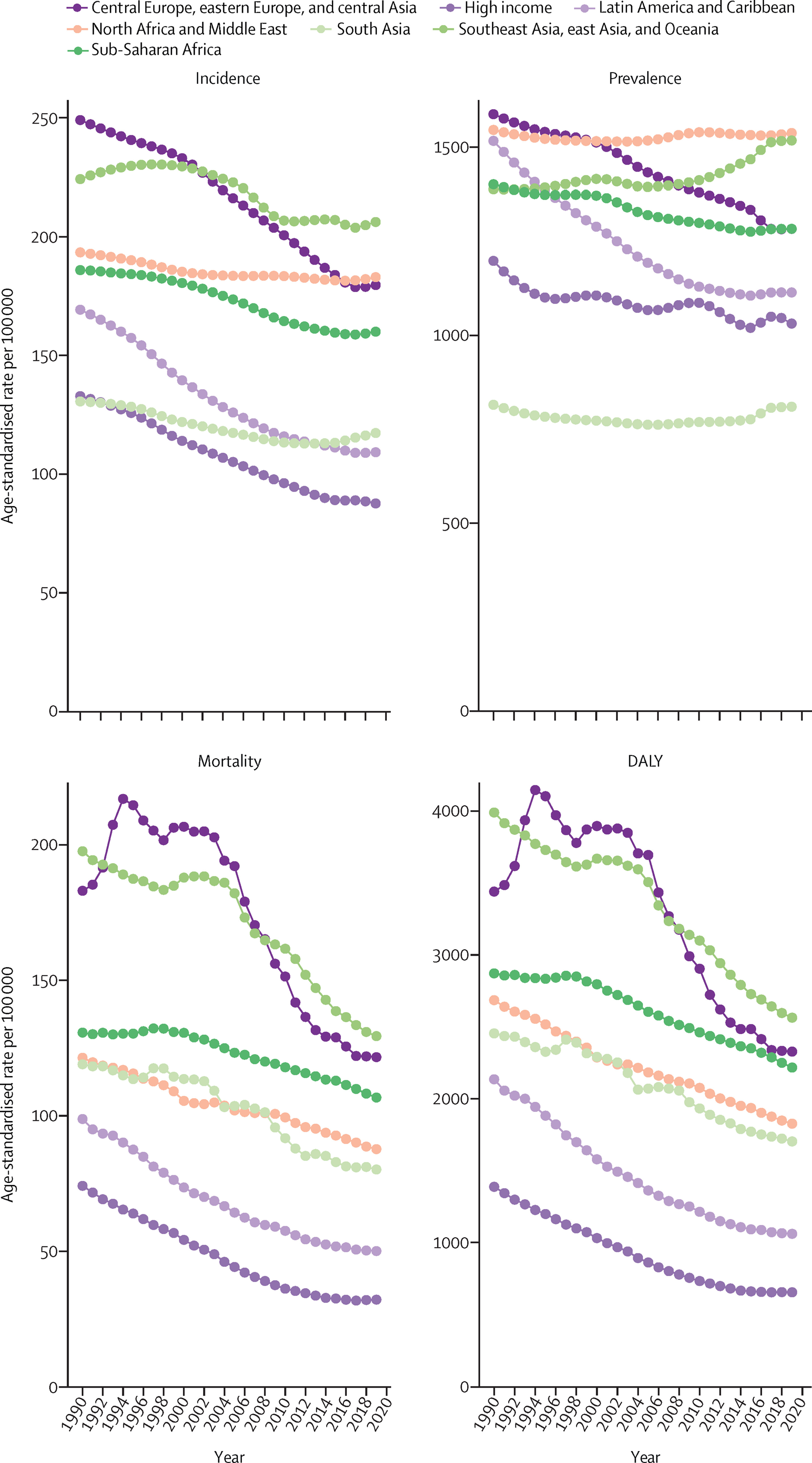}
    \caption{Stroke cases worldwide from 1990 to 2020 \cite{Feigin2021}}
    \label{fig:sample1}
    \hspace{0.05\textwidth}
\end{figure}

Identifying the origin of blood clots that cause strokes presents a significant challenge due to the complicated nature of the underlying pathology. Blood clots can arise from various factors such as coagulopathy, atrial fibrillation, or atherosclerosis, and these factors can interact in intricate ways, making it challenging to pinpoint the cause of the clot.  Additionally, stroke symptoms can differ based on the clot's size and location, further complicating the classification process.\\

Innovative solutions have enabled researchers to create efficient techniques for categorizing the source of blood clots that cause strokes. To achieve this, they employ a range of imaging methods such as MRI and CT, which help identify the clot's position and characteristics. Furthermore, lab evaluations involving blood tests and clotting factor assays, combined with thorough clinical assessments that entail a detailed medical history and physical examination, can offer valuable information on the underlying factors that trigger clot formation. \\

These classification methods offer the significant advantage of guiding treatment decisions. For example, if the clot originates from the heart, anticoagulant therapy may be recommended to prevent future strokes. On the other hand, if the clot is from the carotid artery, a surgical procedure called carotid endarterectomy, which removes plaque from the artery, may be suggested.\\

As technology continues to advance, there has been a growing interest in developing more advanced and precise methods for identifying the origin of stroke blood clots. One of these methods is the use of sophisticated imaging techniques such as diffusion-weighted imaging (DWI), which provides detailed information about the clot's location and extent.  Other imaging techniques like magnetic resonance angiography (MRA) and CT angiography (CTA) can also be employed to identify the blood vessels leading to the clot, thereby allowing for a more precise diagnosis of the clot's source. \\

Laboratory examinations have displayed potential in recognizing the origin of blood clots that cause strokes. Specific biomarkers like D-dimer and brain natriuretic peptide (BNP) can give essential indications about the underlying cause of the blood clot. Genetic testing has also exhibited promise in identifying patients with an elevated risk of developing blood clots, primarily those with a family history of stroke or a recognized genetic disposition. \\

To classify the origin of blood clots that cause strokes, the clinical assessment remains a critical component. An exhaustive medical history can provide significant information about the patient's risk factors for stroke such as hypertension, diabetes, and smoking. However, despite the promising developments in the classification of stroke blood clot origin, numerous obstacles need to be addressed to improve their accuracy and clinical effectiveness.  One of the primary obstacles is the need for more standardized and validated criteria for classification. \\

Additionally, while advanced imaging techniques have demonstrated promise in detecting the source of stroke blood clots, their high cost and limited accessibility may restrict their widespread application in clinical practice.  Despite these challenges, devising more precise and reliable methods for classifying the source of stroke blood clots holds the potential to significantly enhance stroke management and prevention.  By identifying the root cause of the clot, healthcare providers can tailor treatment plans to each patient, resulting in improved outcomes and decreased risk of future strokes.\\

Thus, this study proposes a novel system that can analyze the whole-slide digital pathology image of blood clots from ischemic stroke patients and accurately classify them into Cardioembolic (CE) and Large Artery Atherosclerosis (LAA) and works on par with the pathologists in identifying the origin of stroke blood clots.
    
\section{Related Work}

    Deep learning has made great progress since its inception \cite{8753848}. Neural networks were hampered in the 1980s by a lack of processing power and data. Yet, the advancement of graphics processing units (GPUs) and massive data has made deep neural network training possible \cite{9689188}. This has resulted in advances in image identification, natural language processing, and speech recognition. Deep learning has transformed medical data analysis\cite{9609375}\cite{krishnan2022efficient}  significantly by helping physicians better diagnose numerous diseases and improve overall patient care.\\

    In 2017, Vaswani et al. \cite{vaswani2017attention} presented the transformer architecture as a breakthrough in natural language processing for neural machine translation. By merging self-attention with linear layers, this approach outperformed LSTM-RNN-based methods in various NLP tasks, including neural machine translation. Over time, researchers have proposed various versions and improvements to the transformer architecture across different domains, resulting in superior performance in computer vision \cite{9720737} and natural language processing tasks.\\

    Lie et al. \cite{liu2022swin} developed an enhanced version of the swin transformer architecture by investigating large-scale models in computer vision. The study introduced three methods to address major concerns in the training and use of large-vision models. The strategies comprise a residual-post-norm approach, a log-spaced continuous position bias approach, and a self-supervised pre-training method. Utilizing these methods, the authors trained a 3 billion-parameter Swin Transformer V2 model with high-resolution images. This study achieved new performance records for four representative vision tasks and used 40 times less labeled data and training time than Google's billion-level visual models.\\

    Gupta et al. \cite{9734197} present a comparison of various machine learning models on the task of stroke prediction using different machine learning models, with the AdaBoost, XGBoost, and Random Forest Classifiers having the highest accuracy scores. The study concluded that the RandomForest model achieves the highest accuracy of 97\%.\\

    In a recent study by Rahim et al. \cite{Rahim_Sunyoto_Arief_2022}, the researchers suggest utilizing the Xtreme Gradient Boosting Algorithm as an ensemble learning technique to enhance the accuracy of stroke case predictions. Their results indicate that training data, consisting of 3582 instances, with test data of 1536 instances, produced a 96\% accuracy rate and improved outcomes compared to prior studies.\\

    Tazin et al. \cite{PMID:34868531} have developed multiple machine learning (ML) models to anticipate the probability of a brain stroke happening. Their study revealed that the Random Forest algorithm had the highest accuracy rate, reaching 96\%. The accuracy of the models used in this research surpasses those of previous studies, which indicates the increased reliability of these models.\\

    Darapaneni et al. \cite{10170332} explores deep learning models for classifying stroke-related images. It evaluates CNN architectures like ResNet, VGG, InceptionV3, and DenseNet, highlighting DenseNet with Adam optimizer as most effective. The study also addresses deep learning's "Black box" nature by correlating auto-encoder features with baseline image features. Challenges in processing high-resolution histopathology images are discussed, with ResNet152 and Adam optimizer identified as optimal for this application.\\

    Rao et al. \cite{10176805} propose a novel deep learning approach is utilized to classify the origin of blood clots in ischemic strokes. The model integrates EfficientNet-B0, VGG19, and ResNet-152 architectures to distinguish between cardioembolic and large artery atherosclerotic strokes. Employing the Mayo Clinic - Strip AI dataset, this method demonstrates improved accuracy over existing models, offering significant potential for enhanced diagnosis and management of stroke causes.\\

\section{Methodology}
\subsection{Dataset}

\begin{figure}
    \includegraphics[width=0.50\textwidth, height=40mm]{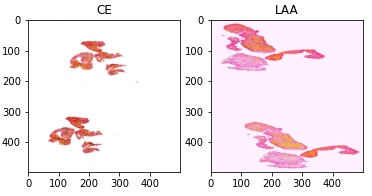}
    \caption{The image on the left is the Cardioembolic blood clot  and the image on the right is the Large Artery Atherosclerosis blood clot }
    \label{fig:classes}
    \hspace{0.05\textwidth}
\end{figure}

\subsubsection{Background classifier dataset}
    To develop a background classifier that could remove empty patches from the whole-slide digital pathology images, we utilized the STRIP AI background clot dataset \cite{strip-ai-background-clot} that was sourced from Kaggle. The dataset comprises a total of 19,998 images, half of which contain cell contents while the other half consists of empty patches without any cell contents.
\subsubsection{Blood clot origin classification dataset}
\begin{figure*}[htbp]
    \centering
    \includegraphics[width=14cm, height=6cm]{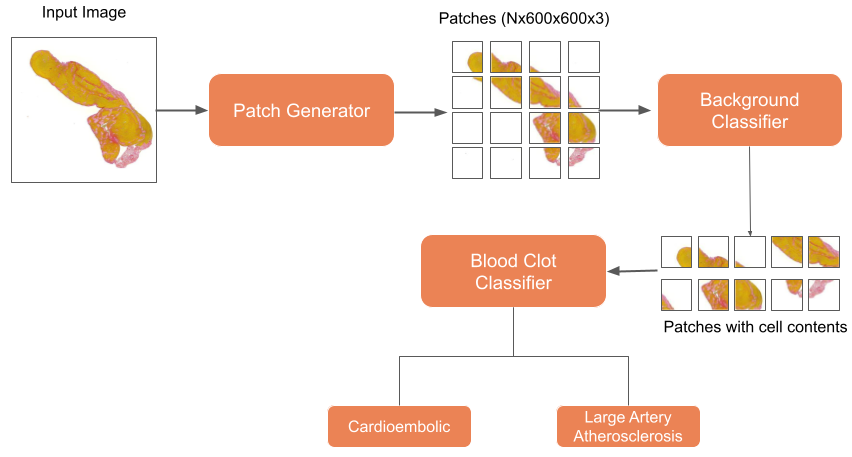}
    \caption{The overall architecture of the proposed system that takes the whole-slide digital pathology image and classifies the blood clot into either Cardioembolic or Large Artery Atherosclerosis.}
    \label{fig:architecture}
\end{figure*}

    To classify between Cardioembolic (CE) and Large Artery Atherosclerosis (LAA) Figure~\ref{fig:classes}, the proposed system was trained on the Mayo Clinic - STRIP AI competition dataset \cite{mayo-clinic-strip-ai}. The dataset comprises whole-slide digital pathology images of blood clots that were obtained from patients who experienced ischemic stroke. The competition aimed to develop a system that could accurately classify the two types of stroke.

    \begin{table}
        \caption{Blood clot origin classification dataset split} 
        \renewcommand*\arraystretch{1.1}
        \normalsize
        \noindent \begin{tabular}{|p{20mm}||p{55mm}|}
        
        \hline
        Split & Number patches with cell contents\\
        \hline
        Train    & 60489 \\
        Validation  & 12962 \\
        Test  & 12963 \\
        \hline
        \textbf{Total}   & \textbf{86414} \\
        \hline
        \end{tabular}
    \end{table}
    

    

\begin{table*}
    \caption{Comparative Performance Analysis of Stroke Blood Clot Origin Classification Models}
    \renewcommand*\arraystretch{1.15}
    \normalsize
    \centering
    \begin{tabular}{|p{32mm}|p{20mm}|p{13mm}|p{13mm}|p{13mm}|p{13mm}|}
    \hline
    Model & Weighted Multi-Class Logarithmic Loss & Accuracy & Precision & Recall & F1-Score\\
    \hline
    \multicolumn{6}{|c|}{\textit{Results from "Image Classification of Ischemic Stroke Blood Clot Origin"}} \\
    \hline
    Best Stacked Model (EfficientNet-B0, VGG19, ResNet-152) & 0.69312 & - & - & - & - \\
    \hline
    \multicolumn{6}{|c|}{\textit{Results from Darapaneni et al. \cite{10170332}}} \\
    \hline
    ResNet152 & - & 0.7484 & - & - & - \\
    CNN & - & \textbf{0.7615} & - & - & - \\
    DenseNet121 & - & 0.6821 & - & - & - \\
    EfficientNet & - & \textbf{0.7615} & - & - & - \\
    \hline
    \multicolumn{6}{|c|}{\textit{Our Results}} \\
    \hline
    ResNet50 & 1.057 & 0.7384 & 0.7384 & 0.7384 & 0.7384\\
    InceptionV3 & 0.5847 & 0.8761 & 0.8761 & 0.8761 & 0.8761\\
    ConvNeXtV2 & 0.4374 & 0.8948 & 0.8948 & 0.8948 & 0.8948\\
    SwinTransformerV2 & 0.3894 & 0.9269 & 0.9269 & 0.9269 & 0.9269\\
    \textbf{PoolFormer} & \textbf{0.4361} & \textbf{0.9345} & \textbf{0.9345} & \textbf{0.9345} & \textbf{0.9345}\\
    \hline
    \end{tabular}
\end{table*}

\subsection{Dataset Preprocessing}
    The data preprocessing step is an essential step in any research project. In this section, we will describe the steps we took to preprocess our data for analysis.

    First, we obtained whole-slide digital pathology images of very large dimensions from the STRIP AI background clot dataset \cite{strip-ai-background-clot}. These images were in the TIFF format, which is a standard format for digital pathology images. To extract patches from these images, we used the OpenSlide \cite{OpenSlide} python library. OpenSlide \cite{OpenSlide} is an open-source library that provides a simple interface for reading digital pathology images. We extracted patches of size 600x600x3 dimensions from the whole-slide digital pathology images and saved them in PNG format.
    
    We then used Otsu's thresholding technique to filter out low-quality patches. The thresholding technique involves calculating the area of the slide contents in the image as well as the total content in the image. If the slide area consists of more than 30\% of the total slide area, the image was kept otherwise it was discarded. This step helped us to remove images with low-quality content that could have affected our analysis negatively.

    Subsequently, we used our previously trained background classifier to remove white images from the dataset. This was done to ensure that the dataset only contains images that are relevant to the study. These steps ensured that the dataset we used for analysis was of high quality and relevance to the study.

    Finally, we used the albumentations \cite{info11020125} library's numerous powerful image augmentation techniques on the image patches. We apply HorizontalFlip, VerticalFlip, RandomAdjustSharpness, Rotate, and ColorJitter transformations are applied on 50\% of the training dataset. The RandomAdjustSharpness transformation was applied with a sharpness factor of 2 and the ColorJitter transformation was applied with brightness=0.2, hue=0.5, and saturation=0.5 values. The Resize transformation which resizes the 600x600x3 patches into 256x256x3 patches and the Normalize transformation which uses the ImageNet mean and standard deviation to normalize the patches are applied on all splits of the dataset.
    
\section{Experiments and Analysis}
\subsection{Experimental Setup}
\subsubsection{Background Classifer}
   At first, we trained a background classifier model on the STRIP AI background clot dataset \cite{strip-ai-background-clot} which contains 9999 images each for both classes. Based on our previous research works \cite{krishnan2023mfaan}, we implemented the transfer learning approach to fine-tune state-of-the-art computer vision models for the task at hand. We wanted to create a lightweight and efficient background classifier and hence we used the MobileNetV3 \cite{howard2019searching} architecture and replaced its classification head with a linear layer that produces the dataset-specific output. This background classifier was trained on images of dimension 128x128x3 to further reduce the inference time. We used the AdamW \cite{loshchilov2019decoupled} optimizer with an initial learning rate of 3e-4.\\
\subsubsection{Blood Clot Classifier}
    The classification of blood clot images into Large Artery Atherosclerosis (LAA) and Cardioembolic (CE) blood clots is crucial in the diagnosis and treatment of stroke. To accomplish this task, we utilized a transfer learning process, which involves fine-tuning pre-trained models on a new dataset. In particular, we used several state-of-the-art models including the swinv2\_tiny\_window16\_256 variant of the SwinTransformerV2 \cite{liu2022swin} architecture, poolformers36 variant of the PoolFormer \cite{yu2022metaformer} architecture, convnext\_small variant of the ConvNeXt \cite{woo2023convnext} architecture, efficientnet\_b3 variant of the EfficientNet \cite{tan2020efficientnet} architecture, resnet50 variant of the ResNet \cite{he2015deep} architecture, and the inceptionv3 \cite{szegedy2015rethinking} architecture, all of which have achieved impressive results in various computer vision tasks. We used the Timm \cite{rw2019timm} python library to get all the above-mentioned pretrained models and fine-tuned them on the ischemic stroke blood clot origin classification task.

    To adapt these pre-trained models to our specific task, we replaced their top layers with a linear layer that produces dataset-specific output. We set the image size to 256 x 256 x 3, which gives a good balance between performance and computational efficiency. Furthermore, to determine the optimal hyperparameters, such as the learning rate and optimizer, we employed the Optuna hyperparameter optimization framework \cite{akiba2019optuna}. This approach involves conducting a systematic search over the hyperparameter space to find the configuration that yields the best performance on the validation set. Through 50 trials using various optimizers and learning rates from 1e-3 to 1e-6, we found that the AdamW optimizer \cite{loshchilov2019decoupled} with an initial learning rate of 3e-4 worked well for all pre-trained models. AdamW optimizer\cite{loshchilov2019decoupled} is an improved version of the popular Adam optimizer that includes weight decay regularization, to optimize the models' parameters during training.
    
    During training, we used two schedulers: StochasticWeightAveraging\cite{izmailov2019averaging} and EarlyStopping. The former computes the average of multiple weight values during training to achieve better generalization, while the latter stops training if the monitored metric did not improve. These schedulers help prevent overfitting and improve the models' generalization performance.
    
    In summary, our approach involved fine-tuning several pre-trained models using transfer learning, optimizing the models' hyperparameters using the Optuna framework \cite{akiba2019optuna}, and using appropriate schedulers to improve the models' performance. The resulting models showed promising results in the classification of blood clot images, which could potentially assist physicians in the diagnosis and treatment of stroke.

\subsection{Evaluation Criteria}
    The Weighted Multi-Class Logarithmic Loss (WMCLL) is used as the loss function to be optimized for this problem. This loss function is a variation of the traditional multi-class logarithmic loss, enhanced by incorporating weights for each class. These weights adjust the contribution of each class to the total loss, emphasizing classes that are underrepresented or of greater importance. WMCLL penalizes incorrect predictions, with the penalty scaled according to the assigned class weights. The loss is calculated using the logarithm of predicted probabilities, ensuring that confident but incorrect predictions are penalized more heavily. 
    \begin{equation}
        WMCLL = -\frac{1}{N} \sum_{i=1}^{N} \sum_{j=1}^{M} w_j \cdot y_{ij} \cdot \log(p_{ij})
    \end{equation}
    where:
    \begin{itemize}
        \item \( N \) is the number of samples.
        \item \( M \) is the number of classes.
        \item \( w_j \) is the weight for class \( j \).
        \item \( y_{ij} \) is 1 if sample \( i \) belongs to class \( j \), else 0.
        \item \( p_{ij} \) is the predicted probability of sample \( i \) for class \( j \).\\\\
    \end{itemize}
    Classification models are commonly evaluated using metrics such as accuracy, precision, recall, and F1 score.
    
    Accuracy is a simple metric that reflects the fraction of properly identified samples among all samples. It is acceptable when the sample sizes in the classes are substantially comparable.
    
    Precision is the percentage of true positive predictions to all positive predictions, and it reflects how frequently the model accurately identifies positive results. When the cost of false positives is significant, this statistic is extremely relevant.

    The fraction of true positive predictions out of all real positive samples is measured as recall, and it reflects how well the model recognizes the positive class. When the cost of false negatives is substantial, this measure is acceptable.

    F1 score is a composite metric that balances precision and recall using their harmonic mean. When the classes are imbalanced, with one class having much more samples than the other, this method is beneficial.

    \begin{equation}
    Accuracy  = \frac{TP+TN}{TP+FP+FN+TN}
    \end{equation}
    \begin{equation}
        Precision = \frac{TP}{TP+FP}
    \end{equation}
    \begin{equation}
        Recall = \frac{TP}{TP+FN}
    \end{equation}
    \begin{equation}
        F1-score = 2 * \frac{Recall * Precision}{Recall + Precision}
    \end{equation}

    In this context, TP (True Positive) denotes the count of accurately classified positive samples, FN (False Negative) represents the quantity of negatively classified samples that were incorrect, TN (True Negative) indicates the number of accurately classified negative samples, and FP (False Positive) refers to the quantity of falsely classified positive samples.

\section{Results}

The summarized results from Table II reveal a competitive landscape in the domain of stroke blood clot origin classification using various deep learning models. The stacked model from Rao et al. \cite{10176805}, combining EfficientNet-B0, VGG19, and ResNet-152, established a baseline with a Logarithmic Loss of 0.69312, although its accuracy and other metrics were not disclosed.\\

In contrast, the ResNet152 model, as reported from the study by Darapaneni et al. \cite{10170332}, achieved an accuracy of 74.84\%, while a standard CNN and EfficientNet tied for the highest accuracy of 76.15\%. DenseNet121 lagged behind with an accuracy of 68.21\%. These results suggest that while ResNet152 is a strong performer, the standard CNN and EfficientNet models are equally competent in this application.\\

Our experimental outcomes further enrich this analysis. The ResNet50 model showed a balanced performance with an accuracy and other metrics uniformly at 73.84\%, while InceptionV3 surpassed it with an accuracy of 87.61\%. More advanced architectures like ConvNeXtV2 and SwinTransformerV2 demonstrated superior performance, with accuracy rates of 89.48\% and 92.69\%, respectively, reflecting their more efficient pattern recognition and generalization capabilities. The standout model in our study was PoolFormer, which not only achieved an excellent Logarithmic Loss of 0.4361 but also recorded the highest accuracy, precision, recall, and F1-score at 93.45\%. This performance highlights the PoolFormer's adeptness at handling the complexity of medical image classification and suggests its suitability for practical deployment in medical diagnostics.

\section{Conclusion}
Finally, the proposed stroke blood clot origin detection system is a valuable tool that can help doctors accurately determine the origin of a blood clot-caused stroke. The proposed system employs deep learning algorithms to give doctors a dependable and quick method of diagnosing the type of stroke and determining the best treatment approach. This system has the potential to save lives, save healthcare costs, and improve overall care quality. Overall, the proposed stroke blood clot origin detection system offers a substantial advancement in stroke diagnosis and therapy. It has the potential, with more development and refining, to transform how doctors approach stroke care, ultimately saving the lives of patients all around the world.

\section{Future Work}    
Although the proposed system achieves higher accuracies in identifying the origin of blood clots, there is still room for improvement. The following are some of the approaches to improve the proposed system's performance.
\begin{enumerate}
\item Improving the accuracy of the classification algorithm is an essential topic of future effort.  This could include employing more complex deep learning algorithms or incorporating additional data sources to increase the model's accuracy. 

\item To boost the precision and effectiveness of the categorization, patient-specific data could be incorporated into the algorithm. Data such as age, gender, medical history, and other parameters could be used to better forecast the origin of a blood clot.
\end{enumerate}

\balance
\printbibliography

@Article{info11020125,
    AUTHOR = {Buslaev, Alexander and Iglovikov, Vladimir I. and Khvedchenya, Eugene and Parinov, Alex and Druzhinin, Mikhail and Kalinin, Alexandr A.},
    TITLE = {Albumentations: Fast and Flexible Image Augmentations},
    JOURNAL = {Information},
    VOLUME = {11},
    YEAR = {2020},
    NUMBER = {2},
    ARTICLE-NUMBER = {125},
    URL = {https://www.mdpi.com/2078-2489/11/2/125},
    ISSN = {2078-2489},
    DOI = {10.3390/info11020125}
}

@article{krishnan2022efficient,
  title={Efficient Super-Resolution For Chest X-rays},
  author={Krishnan, Karthik Sivarama and Krishnan, Koushik Sivarama},
  journal={Acta Scientific COMPUTER SCIENCES Volume},
  volume={4},
  number={6},
  year={2022}
}

@misc{mayo-clinic-strip-ai,
    author={Ashley Chow and Barbaros and Ryan Holbrook and Sobhi Jabal and VikashGupta},
    title={Mayo Clinic - STRIP AI},
    publisher={Kaggle},
    year={2022},
    url={https://kaggle.com/competitions/mayo-clinic-strip-ai}
}

@misc{strip-ai-background-clot,
    author={alejopaullier},
    title={strip-ai-background-clot},
    publisher={Kaggle},
    year={2022},
    url={https://www.kaggle.com/datasets/alejopaullier/strip-ai-background-clot}
}

@article{Rahim_Sunyoto_Arief_2022, 
title={Stroke Prediction Using Machine Learning Method with Extreme Gradient Boosting Algorithm}, 
volume={21},
url={https://journal.universitasbumigora.ac.id/index.php/matrik/article/view/1666}, 
DOI={https://doi.org/10.30812/matrik.v21i3.1666}, 
number={3}, 
journal={MATRIK : Jurnal Manajemen, Teknik Informatika dan Rekayasa Komputer}, 
author={Rahim, Abd Mizwar A and Sunyoto, Andi and Arief, Muhammad Rudyanto}, 
year={2022}, 
month={Jul.}, 
pages={595-606}
}

@article {PMID:34868531,
	Title = {Stroke Disease Detection and Prediction Using Robust Learning Approaches},
	Author = {Tazin, Tahia and Alam, Md Nur and Dola, Nahian Nakiba and Bari, Mohammad Sajibul and Bourouis, Sami and Monirujjaman Khan, Mohammad},
	DOI = {10.1155/2021/7633381},
	Volume = {2021},
	Year = {2021},
	Journal = {Journal of healthcare engineering},
	ISSN = {2040-2295},
	Pages = {7633381},
	URL = {https://europepmc.org/articles/PMC8641997},
}

@INPROCEEDINGS{9734197,
  author={Gupta, Saumya and Raheja, Supriya},
  booktitle={2022 12th International Conference on Cloud Computing, Data Science and Engineering (Confluence)}, 
  title={Stroke Prediction using Machine Learning Methods}, 
  year={2022},
  volume={},
  number={},
  pages={553-558},
  doi={10.1109/Confluence52989.2022.9734197}}

@Article{Feigin2021,
author={Feigin, Valery L.
and Stark, Benjamin A.
and Johnson, Catherine Owens
and Roth, Gregory A.
and Bisignano, Catherine
and Abady, Gdiom Gebreheat
and Abbasifard, Mitra
and Abbasi-Kangevari, Mohsen
and Abd-Allah, Foad
and Abedi, Vida
and Abualhasan, Ahmed
and Abu-Rmeileh, Niveen ME
and Abushouk, Abdelrahman I.
and Adebayo, Oladimeji M.
and Agarwal, Gina
and Agasthi, Pradyumna
and Ahinkorah, Bright Opoku
and Ahmad, Sohail
and Ahmadi, Sepideh
and Ahmed Salih, Yusra
and Aji, Budi
and Akbarpour, Samaneh
and Akinyemi, Rufus Olusola
and Al Hamad, Hanadi
and Alahdab, Fares
and Alif, Sheikh Mohammad
and Alipour, Vahid
and Aljunid, Syed Mohamed
and Almustanyir, Sami
and Al-Raddadi, Rajaa M.
and Al-Shahi Salman, Rustam
and Alvis-Guzman, Nelson
and Ancuceanu, Robert
and Anderlini, Deanna
and Anderson, Jason A.
and Ansar, Adnan
and Antonazzo, Ippazio Cosimo
and Arabloo, Jalal
and {\"A}rnl{\"o}v, Johan
and Artanti, Kurnia Dwi
and Aryan, Zahra
and Asgari, Samaneh
and Ashraf, Tahira
and Athar, Mohammad
and Atreya, Alok
and Ausloos, Marcel
and Baig, Atif Amin
and Baltatu, Ovidiu Constantin
and Banach, Maciej
and Barboza, Miguel A.
and Barker-Collo, Suzanne Lyn
and B{\"a}rnighausen, Till Winfried
and Barone, Mark Thomaz Ugliara
and Basu, Sanjay
and Bazmandegan, Gholamreza
and Beghi, Ettore
and Beheshti, Mahya
and B{\'e}jot, Yannick
and Bell, Arielle Wilder
and Bennett, Derrick A.
and Bensenor, Isabela M.
and Bezabhe, Woldesellassie Mequanint
and Bezabih, Yihienew Mequanint
and Bhagavathula, Akshaya Srikanth
and Bhardwaj, Pankaj
and Bhattacharyya, Krittika
and Bijani, Ali
and Bikbov, Boris
and Birhanu, Mulugeta M.
and Boloor, Archith
and Bonny, Aime
and Brauer, Michael
and Brenner, Hermann
and Bryazka, Dana
and Butt, Zahid A.
and Caetano dos Santos, Florentino Luciano
and Campos-Nonato, Ismael R.
and Cantu-Brito, Carlos
and Carrero, Juan J.
and Casta{\~{n}}eda-Orjuela, Carlos A.
and Catapano, Alberico L.
and Chakraborty, Promit Ananyo
and Charan, Jaykaran
and Choudhari, Sonali Gajanan
and Chowdhury, Enayet Karim
and Chu, Dinh-Toi
and Chung, Sheng-Chia
and Colozza, David
and Costa, Vera Marisa
and Costanzo, Simona
and Criqui, Michael H.
and Dadras, Omid
and Dagnew, Baye
and Dai, Xiaochen
and Dalal, Koustuv
and Damasceno, Albertino Antonio Moura
and D'Amico, Emanuele
and Dandona, Lalit
and Dandona, Rakhi
and Darega Gela, Jiregna
and Davletov, Kairat
and De la Cruz-G{\'o}ngora, Vanessa
and Desai, Rupak
and Dhamnetiya, Deepak
and Dharmaratne, Samath Dhamminda
and Dhimal, Mandira Lamichhane
and Dhimal, Meghnath
and Diaz, Daniel
and Dichgans, Martin
and Dokova, Klara
and Doshi, Rajkumar
and Douiri, Abdel
and Duncan, Bruce B.
and Eftekharzadeh, Sahar
and Ekholuenetale, Michael
and El Nahas, Nevine
and Elgendy, Islam Y.
and Elhadi, Muhammed
and El-Jaafary, Shaimaa I.
and Endres, Matthias
and Endries, Aman Yesuf
and Erku, Daniel Asfaw
and Faraon, Emerito Jose A.
and Farooque, Umar
and Farzadfar, Farshad
and Feroze, Abdullah Hamid
and Filip, Irina
and Fischer, Florian
and Flood, David
and Gad, Mohamed M.
and Gaidhane, Shilpa
and Ghanei Gheshlagh, Reza
and Ghashghaee, Ahmad
and Ghith, Nermin
and Ghozali, Ghozali
and Ghozy, Sherief
and Gialluisi, Alessandro
and Giampaoli, Simona
and Gilani, Syed Amir
and Gill, Paramjit Singh
and Gnedovskaya, Elena V.
and Golechha, Mahaveer
and Goulart, Alessandra C.
and Guo, Yuming
and Gupta, Rajeev
and Gupta, Veer Bala
and Gupta, Vivek Kumar
and Gyanwali, Pradip
and Hafezi-Nejad, Nima
and Hamidi, Samer
and Hanif, Asif
and Hankey, Graeme J.
and Hargono, Arief
and Hashi, Abdiwahab
and Hassan, Treska S.
and Hassen, Hamid Yimam
and Havmoeller, Rasmus J.
and Hay, Simon I.
and Hayat, Khezar
and Hegazy, Mohamed I.
and Herteliu, Claudiu
and Holla, Ramesh
and Hostiuc, Sorin
and Househ, Mowafa
and Huang, Junjie
and Humayun, Ayesha
and Hwang, Bing-Fang
and Iacoviello, Licia
and Iavicoli, Ivo
and Ibitoye, Segun Emmanuel
and Ilesanmi, Olayinka Stephen
and Ilic, Irena M.
and Ilic, Milena D.
and Iqbal, Usman
and Irvani, Seyed Sina Naghibi
and Islam, Sheikh Mohammed Shariful
and Ismail, Nahlah Elkudssiah
and Iso, Hiroyasu
and Isola, Gaetano
and Iwagami, Masao
and Jacob, Louis
and Jain, Vardhmaan
and Jang, Sung-In
and Jayapal, Sathish Kumar
and Jayaram, Shubha
and Jayawardena, Ranil
and Jeemon, Panniyammakal
and Jha, Ravi Prakash
and Johnson, Walter D.
and Jonas, Jost B.
and Joseph, Nitin
and Jozwiak, Jacek Jerzy
and J{\"u}risson, Mikk
and Kalani, Rizwan
and Kalhor, Rohollah
and Kalkonde, Yogeshwar
and Kamath, Ashwin
and Kamiab, Zahra
and Kanchan, Tanuj
and Kandel, Himal
and Karch, Andr{\'e}
and Katoto, Patrick DMC
and Kayode, Gbenga A.
and Keshavarz, Pedram
and Khader, Yousef Saleh
and Khan, Ejaz Ahmad
and Khan, Imteyaz A.
and Khan, Maseer
and Khan, Moien AB
and Khatib, Mahalaqua Nazli
and Khubchandani, Jagdish
and Kim, Gyu Ri
and Kim, Min Seo
and Kim, Yun Jin
and Kisa, Adnan
and Kisa, Sezer
and Kivim{\"a}ki, Mika
and Kolte, Dhaval
and Koolivand, Ali
and Koulmane Laxminarayana, Sindhura Lakshmi
and Koyanagi, Ai
and Krishan, Kewal
and Krishnamoorthy, Vijay
and Krishnamurthi, Rita V.
and Kumar, G. Anil
and Kusuma, Dian
and La Vecchia, Carlo
and Lacey, Ben
and Lak, Hassan Mehmood
and Lallukka, Tea
and Lasrado, Savita
and Lavados, Pablo M.
and Leonardi, Matilde
and Li, Bingyu
and Li, Shanshan
and Lin, Hualiang
and Lin, Ro-Ting
and Liu, Xuefeng
and Lo, Warren David
and Lorkowski, Stefan
and Lucchetti, Giancarlo
and Lutzky Saute, Ricardo
and Magdy Abd El Razek, Hassan
and Magnani, Francesca Giulia
and Mahajan, Preetam Bhalchandra
and Majeed, Azeem
and Makki, Alaa
and Malekzadeh, Reza
and Malik, Ahmad Azam
and Manafi, Navid
and Mansournia, Mohammad Ali
and Mantovani, Lorenzo Giovanni
and Martini, Santi
and Mazzaglia, Giampiero
and Mehndiratta, Man Mohan
and Menezes, Ritesh G.
and Meretoja, Atte
and Mersha, Amanual Getnet
and Miao Jonasson, Junmei
and Miazgowski, Bartosz
and Miazgowski, Tomasz
and Michalek, Irmina Maria
and Mirrakhimov, Erkin M.
and Mohammad, Yousef
and Mohammadian-Hafshejani, Abdollah
and Mohammed, Shafiu
and Mokdad, Ali H.
and Mokhayeri, Yaser
and Molokhia, Mariam
and Moni, Mohammad Ali
and Montasir, Ahmed Al
and Moradzadeh, Rahmatollah
and Morawska, Lidia
and Morze, Jakub
and Muruet, Walter
and Musa, Kamarul Imran
and Nagarajan, Ahamarshan Jayaraman
and Naghavi, Mohsen
and Narasimha Swamy, Sreenivas
and Nascimento, Bruno Ramos
and Negoi, Ruxandra Irina
and Neupane Kandel, Sandhya
and Nguyen, Trang Huyen
and Norrving, Bo
and Noubiap, Jean Jacques
and Nwatah, Vincent Ebuka
and Oancea, Bogdan
and Odukoya, Oluwakemi Ololade
and Olagunju, Andrew T.
and Orru, Hans
and Owolabi, Mayowa O.
and Padubidri, Jagadish Rao
and Pana, Adrian
and Parekh, Tarang
and Park, Eun-Cheol
and Pashazadeh Kan, Fatemeh
and Pathak, Mona
and Peres, Mario F. P.
and Perianayagam, Arokiasamy
and Pham, Truong-Minh
and Piradov, Michael A.
and Podder, Vivek
and Polinder, Suzanne
and Postma, Maarten J.
and Pourshams, Akram
and Radfar, Amir
and Rafiei, Alireza
and Raggi, Alberto
and Rahim, Fakher
and Rahimi-Movaghar, Vafa
and Rahman, Mosiur
and Rahman, Muhammad Aziz
and Rahmani, Amir Masoud
and Rajai, Nazanin
and Ranasinghe, Priyanga
and Rao, Chythra R.
and Rao, Sowmya J.
and Rathi, Priya
and Rawaf, David Laith
and Rawaf, Salman
and Reitsma, Marissa B.
and Renjith, Vishnu
and Renzaho, Andre M. N.
and Rezapour, Aziz
and Rodriguez, Jefferson Antonio Buendia
and Roever, Leonardo
and Romoli, Michele
and Rynkiewicz, Andrzej
and Sacco, Simona
and Sadeghi, Masoumeh
and Saeedi Moghaddam, Sahar
and Sahebkar, Amirhossein
and Saif-Ur-Rahman, K. M.
and Salah, Rehab
and Samaei, Mehrnoosh
and Samy, Abdallah M.
and Santos, Itamar S.
and Santric-Milicevic, Milena M.
and Sarrafzadegan, Nizal
and Sathian, Brijesh
and Sattin, Davide
and Schiavolin, Silvia
and Schlaich, Markus P.
and Schmidt, Maria In{\^e}s
and Schutte, Aletta Elisabeth
and Sepanlou, Sadaf G.
and Seylani, Allen
and Sha, Feng
and Shahabi, Saeed
and Shaikh, Masood Ali
and Shannawaz, Mohammed
and Shawon, Md Shajedur Rahman
and Sheikh, Aziz
and Sheikhbahaei, Sara
and Shibuya, Kenji
and Siabani, Soraya
and Silva, Diego Augusto Santos
and Singh, Jasvinder A.
and Singh, Jitendra Kumar
and Skryabin, Valentin Yurievich
and Skryabina, Anna Aleksandrovna
and Sobaih, Badr Hasan
and Stortecky, Stefan
and Stranges, Saverio
and Tadesse, Eyayou Girma
and Tarigan, Ingan Ukur
and Temsah, Mohamad-Hani
and Teuschl, Yvonne
and Thrift, Amanda G.
and Tonelli, Marcello
and Tovani-Palone, Marcos Roberto
and Tran, Bach Xuan
and Tripathi, Manjari
and Tsegaye, Gebiyaw Wudie
and Ullah, Anayat
and Unim, Brigid
and Unnikrishnan, Bhaskaran
and Vakilian, Alireza
and Valadan Tahbaz, Sahel
and Vasankari, Tommi Juhani
and Venketasubramanian, Narayanaswamy
and Vervoort, Dominique
and Vo, Bay
and Volovici, Victor
and Vosoughi, Kia
and Vu, Giang Thu
and Vu, Linh Gia
and Wafa, Hatem A.
and Waheed, Yasir
and Wang, Yanzhong
and Wijeratne, Tissa
and Winkler, Andrea Sylvia
and Wolfe, Charles D. A.
and Woodward, Mark
and Wu, Jason H.
and Wulf Hanson, Sarah
and Xu, Xiaoyue
and Yadav, Lalit
and Yadollahpour, Ali
and Yahyazadeh Jabbari, Seyed Hossein
and Yamagishi, Kazumasa
and Yatsuya, Hiroshi
and Yonemoto, Naohiro
and Yu, Chuanhua
and Yunusa, Ismaeel
and Zaman, Muhammed Shahriar
and Zaman, Sojib Bin
and Zamanian, Maryam
and Zand, Ramin
and Zandifar, Alireza
and Zastrozhin, Mikhail Sergeevich
and Zastrozhina, Anasthasia
and Zhang, Yunquan
and Zhang, Zhi-Jiang
and Zhong, Chenwen
and Zuniga, Yves Miel H.
and Murray, Christopher J. L.},
title={Global, regional, and national burden of stroke and its risk factors, 1990{\&}{\#}x2013;2019: a systematic analysis for the Global Burden of Disease Study 2019},
journal={The Lancet Neurology},
year={2021},
month={Oct},
day={01},
publisher={Elsevier},
volume={20},
number={10},
pages={795-820},
issn={1474-4422},
doi={10.1016/S1474-4422(21)00252-0},
url={https://doi.org/10.1016/S1474-4422(21)00252-0}
}

@article{OpenSlide,
author = {Goode, Adam and Gilbert, Benjamin and Harkes, Jan and Jukic, Drazen and Satyanarayanan, Mahadev},
year = {2013},
month = {09},
pages = {27},
title = {OpenSlide: A vendor-neutral software foundation for digital pathology},
volume = {4},
journal = {Journal of pathology informatics},
doi = {10.4103/2153-3539.119005}
}

@misc{loshchilov2019decoupled,
      title={Decoupled Weight Decay Regularization}, 
      author={Ilya Loshchilov and Frank Hutter},
      year={2019},
      eprint={1711.05101},
      archivePrefix={arXiv},
      primaryClass={cs.LG}
}

@misc{howard2019searching,
      title={Searching for MobileNetV3}, 
      author={Andrew Howard and Mark Sandler and Grace Chu and Liang-Chieh Chen and Bo Chen and Mingxing Tan and Weijun Wang and Yukun Zhu and Ruoming Pang and Vijay Vasudevan and Quoc V. Le and Hartwig Adam},
      year={2019},
      eprint={1905.02244},
      archivePrefix={arXiv},
      primaryClass={cs.CV}
}

@misc{akiba2019optuna,
      title={Optuna: A Next-generation Hyperparameter Optimization Framework}, 
      author={Takuya Akiba and Shotaro Sano and Toshihiko Yanase and Takeru Ohta and Masanori Koyama},
      year={2019},
      eprint={1907.10902},
      archivePrefix={arXiv},
      primaryClass={cs.LG}
}

@misc{yu2022metaformer,
      title={MetaFormer Is Actually What You Need for Vision}, 
      author={Weihao Yu and Mi Luo and Pan Zhou and Chenyang Si and Yichen Zhou and Xinchao Wang and Jiashi Feng and Shuicheng Yan},
      year={2022},
      eprint={2111.11418},
      archivePrefix={arXiv},
      primaryClass={cs.CV}
}

@misc{liu2022swin,
      title={Swin Transformer V2: Scaling Up Capacity and Resolution}, 
      author={Ze Liu and Han Hu and Yutong Lin and Zhuliang Yao and Zhenda Xie and Yixuan Wei and Jia Ning and Yue Cao and Zheng Zhang and Li Dong and Furu Wei and Baining Guo},
      year={2022},
      eprint={2111.09883},
      archivePrefix={arXiv},
      primaryClass={cs.CV}
}

@misc{tan2020efficientnet,
      title={EfficientNet: Rethinking Model Scaling for Convolutional Neural Networks}, 
      author={Mingxing Tan and Quoc V. Le},
      year={2020},
      eprint={1905.11946},
      archivePrefix={arXiv},
      primaryClass={cs.LG}
}

@INPROCEEDINGS{8753848,
  author={Krishnan, Karthik Sivarama and Sahin, Ferat},
  booktitle={2019 14th Annual Conference System of Systems Engineering (SoSE)}, 
  title={ORBDeepOdometry - A Feature-Based Deep Learning Approach to Monocular Visual Odometry}, 
  year={2019},
  volume={},
  number={},
  pages={296-301},
  doi={10.1109/SYSOSE.2019.8753848}}

@INPROCEEDINGS{9689188,
  author={Krishnan, Koushik Sivarama and Krishnan, Karthik Sivarama},
  booktitle={2021 International Conference on Intelligent Cybernetics Technology and Applications (ICICyTA)}, 
  title={SwiftSRGAN - Rethinking Super-Resolution for Efficient and Real-time Inference}, 
  year={2021},
  volume={},
  number={},
  pages={46-51},
  doi={10.1109/ICICyTA53712.2021.9689188}}

@INPROCEEDINGS{9720737,
  author={Krishnan, Karthik Sivarama and Krishnan, Koushik Sivarama},
  booktitle={2022 IEEE 12th Annual Computing and Communication Workshop and Conference (CCWC)}, 
  title={Benchmarking Conventional Vision Models on Neuromorphic Fall Detection and Action Recognition Dataset}, 
  year={2022},
  volume={},
  number={},
  pages={0518-0523},
  doi={10.1109/CCWC54503.2022.9720737}}

@misc{he2015deep,
      title={Deep Residual Learning for Image Recognition}, 
      author={Kaiming He and Xiangyu Zhang and Shaoqing Ren and Jian Sun},
      year={2015},
      eprint={1512.03385},
      archivePrefix={arXiv},
      primaryClass={cs.CV}
}

@misc{szegedy2015rethinking,
      title={Rethinking the Inception Architecture for Computer Vision}, 
      author={Christian Szegedy and Vincent Vanhoucke and Sergey Ioffe and Jonathon Shlens and Zbigniew Wojna},
      year={2015},
      eprint={1512.00567},
      archivePrefix={arXiv},
      primaryClass={cs.CV}
}

@INPROCEEDINGS{10170332,
  author={Darapaneni, Narayana and Sudha, B.G and Reddy, Anwesh and Abdul Karim, Ab and Marothu, Dhanalakshmi and Kulkarni, Shirish and Menon, Deepak Das},
  booktitle={2023 International Conference on Signal Processing, Computation, Electronics, Power and Telecommunication (IConSCEPT)}, 
  title={Image Classification of Stroke Blood Clot Origin}, 
  year={2023},
  volume={},
  number={},
  pages={1-6},
  doi={10.1109/IConSCEPT57958.2023.10170332}}

@misc{krishnan2023mfaan,
      title={MFAAN: Unveiling Audio Deepfakes with a Multi-Feature Authenticity Network}, 
      author={Karthik Sivarama Krishnan and Koushik Sivarama Krishnan},
      year={2023},
      eprint={2311.03509},
      archivePrefix={arXiv},
      primaryClass={cs.SD}
}

@INPROCEEDINGS{10176805,
  author={Rao, Masabathula V S Raghavendra and Puligundla, Surya and Ekkala, Nikhil Sai and Chebrolu, Srilatha},
  booktitle={2023 Third International Conference on Secure Cyber Computing and Communication (ICSCCC)}, 
  title={Image Classification of Ischemic Stroke Blood Clot Origin using Stacked EfficientNet-B0, VGG19 and ResNet-152}, 
  year={2023},
  volume={},
  number={},
  pages={304-310},
  doi={10.1109/ICSCCC58608.2023.10176805}}

@misc{woo2023convnext,
      title={ConvNeXt V2: Co-designing and Scaling ConvNets with Masked Autoencoders}, 
      author={Sanghyun Woo and Shoubhik Debnath and Ronghang Hu and Xinlei Chen and Zhuang Liu and In So Kweon and Saining Xie},
      year={2023},
      eprint={2301.00808},
      archivePrefix={arXiv},
      primaryClass={cs.CV}
}

@misc{rw2019timm,
  author = {Ross Wightman},
  title = {PyTorch Image Models},
  year = {2019},
  publisher = {GitHub},
  journal = {GitHub repository},
  doi = {10.5281/zenodo.4414861},
  howpublished = {\url{https://github.com/rwightman/pytorch-image-models}}
}

@misc{izmailov2019averaging,
      title={Averaging Weights Leads to Wider Optima and Better Generalization}, 
      author={Pavel Izmailov and Dmitrii Podoprikhin and Timur Garipov and Dmitry Vetrov and Andrew Gordon Wilson},
      year={2019},
      eprint={1803.05407},
      archivePrefix={arXiv},
      primaryClass={cs.LG}
}

@INPROCEEDINGS{9609375,
  author={Krishnan, Koushik Sivarama and Krishnan, Karthik Sivarama},
  booktitle={2021 6th International Conference on Signal Processing, Computing and Control (ISPCC)}, 
  title={Vision Transformer based COVID-19 Detection using Chest X-rays}, 
  year={2021},
  volume={},
  number={},
  pages={644-648},
  doi={10.1109/ISPCC53510.2021.9609375}}

@misc{worldstrokeorg,
  title = "Learn About Stroke",
  howpublished = {\url{https://www.world-stroke.org/world-stroke-day-campaign/why-stroke-matters/learn-about-stroke}},
}

@misc{vaswani2017attention,
      title={Attention Is All You Need}, 
      author={Ashish Vaswani and Noam Shazeer and Niki Parmar and Jakob Uszkoreit and Llion Jones and Aidan N. Gomez and Lukasz Kaiser and Illia Polosukhin},
      year={2017},
      eprint={1706.03762},
      archivePrefix={arXiv},
      primaryClass={cs.CL}
}
\end{document}